\documentclass{article}
\usepackage{spconf,amsmath,graphicx,cite}

\usepackage{color}
\usepackage{enumitem}
\usepackage{hyperref}
\usepackage{multirow}
\usepackage{etoolbox,siunitx}
\robustify\bfseries
\usepackage{booktabs}
\usepackage{balance}
\usepackage{amssymb}
\usepackage{bm}

\DeclareDocumentCommand \Lclip{ m g }{{\mathcal{L}_{\operatorname{c-#1} \IfNoValueF {#2} {,{#2}}}}}
\DeclareDocumentCommand \Lframe{ m g }{{\mathcal{L}_{\operatorname{f-#1} \IfNoValueF {#2} {,{#2}}}}}
\DeclareDocumentCommand \Lharm{ m g }{{\mathcal{L}_{\operatorname{h-#1} \IfNoValueF {#2} {,{#2}}}}}

\title{Transcription Is All You Need:\\Learning to Separate Musical Mixtures with Score as Supervision} %
\name{Yun-Ning Hung$^{1,2}$,
      Gordon Wichern$^{1}$,
      Jonathan Le Roux,$^{1}$\thanks{This work was performed while Y.~Hung was an intern at MERL.}}
\address{$^1$Mitsubishi Electric Research Laboratories (MERL), Cambridge, MA, USA\\$^{2}$Center for Music Technology, Georgia Institute of Technology, Atlanta, GA, USA\\  
         {\small\texttt{amyhung@gatech.edu, \{wichern,leroux\}@merl.com}}}
\begin{document}
\ninept
\maketitle
\begin{abstract}
Most music source separation systems require large collections of isolated sources for training, which can be difficult to obtain. In this work, we use musical scores, which are comparatively easy to obtain, as a weak label for training a source separation system. In contrast with previous score-informed separation approaches, our system does not require isolated sources, and score is used only as a training target, not required for inference. Our model consists of a separator that outputs a time-frequency mask for each instrument, and a transcriptor that acts as a critic, providing both temporal and frequency supervision to guide the learning of the separator. A harmonic mask constraint is introduced as another way of leveraging score information during training, and we propose two novel adversarial losses for additional fine-tuning of both the transcriptor and the separator. Results demonstrate that using score information outperforms temporal weak-labels, and adversarial structures lead to further improvements in both separation and transcription performance.

\end{abstract}
\section{Introduction}
\label{sec:intro}
Music source separation has long been an important task for music information retrieval, and recent advances in deep neural networks, have led to dramatic performance improvements. Most current methods rely on supervised learning, which requires a dataset including separated tracks, or stems for each instrument. However, copyright issues prevent wide availability of stems for most commercial music, and open-source datasets suffer from various drawbacks. 
For example, MUSDB \cite{musdb18} has a limited number of instruments, MIR-1K \cite{Hsu2010OnTI} only contains short clips of music, and MedleyDB \cite{bittner2014medleydb} features an unbalanced amount of instrument categories. Most importantly, compared to large-scale datasets such as AudioSet \cite{gemmeke2017audio} and The Million Song dataset \cite{bertin2011million}, which are used in audio or music classification, source separation datasets are relatively small.

In contrast, it is more practical to obtain a musical score for a track than its isolated stems. Trained musicians are able to transcribe either part of the instruments or the whole song, which contributes to a large number of accessible scores. For example, the Musescore \cite{hung2019multitask} and Lakh MIDI \cite{raffel2016learning} datasets were both collected from online forums, while the SIMSSA project \cite{calvo2018deep} provides the ability to generate digital scores from existing classical sheet music.

Based on this data advantage, in this work, we investigate whether score information alone can be used as a weak label to train a music source separation system without access to isolated stems. %
Many previous works have studied incorporating score information into music source separation, either as an additional training target to learn musically relevant source separation models~\cite{manilow2020simultaneous}, or as part of ``score-informed'' source separation approaches~\cite{Miron2016ScoreInformedSS, Miron2017MonauralSS, ewert2017structured, slizovskaia2019end, schulze2019weakly}, where scores are used as an additional input feature or conditioning mechanism.
However, all these methods are still trained in a supervised manner, and most need the score during inference as well. Different from these works, our proposed approach does not need to use isolated sources during training. The model directly learns from musical scores, and only needs the music mixture during inference. 

In addition to musical score, other types of weak labels have recently been used for sound event separation. For example, \cite{karamatli2019audio, pishdadian2020finding} introduce model structures that only need class labels for training a source separation system, rather than labels on each time-frequency bin. Instead of using ground truth class labels, \cite{Tzinis_ICASSP2020, kong2020source} propose to combine sound event detection systems with source separation systems to achieve large-scale sound separation. 
Visual information is leveraged to separate sound in~\cite{gao2019co, zhao2018sound}, where video features are available at both training and inference time.
Generative adversarial network architectures have also been proposed to learn the distribution of isolated sources \cite{zhang2017weakly, stoller2018adversarial}, without requiring %
isolated sources associated with an input mixture. 
Without observing the target source in isolation, the models in~\cite{stowell2015denoising, michelashvili2019semi} learn to separate from corrupted target observations and regions of isolated interference. 

Our problem definition is similar to that of Pishdadian et al~\cite{pishdadian2020finding}, in that only weak labels are available as training targets, not the strong labels obtained from isolated sources.  However, the weak labels we consider here are musical scores, rather than temporal labels of source activity. While it may sound more restrictive, this is a realistic assumption for music separation. Furthermore, using scores allows us to cope with several drawbacks that may occur when extending the approach in \cite{pishdadian2020finding}, which relies on supervision by a classifier (pre-trained on audio mixtures and corresponding source activity labels), to the music domain.
First, since only temporal information regarding source presence is provided to the classifier, harmonic sounds (e.g., sirens or instruments) are not well separated, as temporal information alone is not enough to isolate upper harmonics. Second, the model does not learn to separate well when two sounds consistently appear together, a scenario that often occurs in music. For example, bass and drums often consistently play through entire songs. To solve the above problems, we propose using a transcriptor (pre-trained on mixtures and their musical scores) to provide both temporal and frequency information when training the separator. Moreover, we incorporate harmonic masks generated form the score to better model harmonic structure, and present two novel adversarial fine-tuning strategies to improve the ability of the separator to learn from weak labels.  %

\section{Training with transcription as supervision}

\begin{figure*}[t]
    \centering
    \includegraphics[width =
    2\columnwidth]{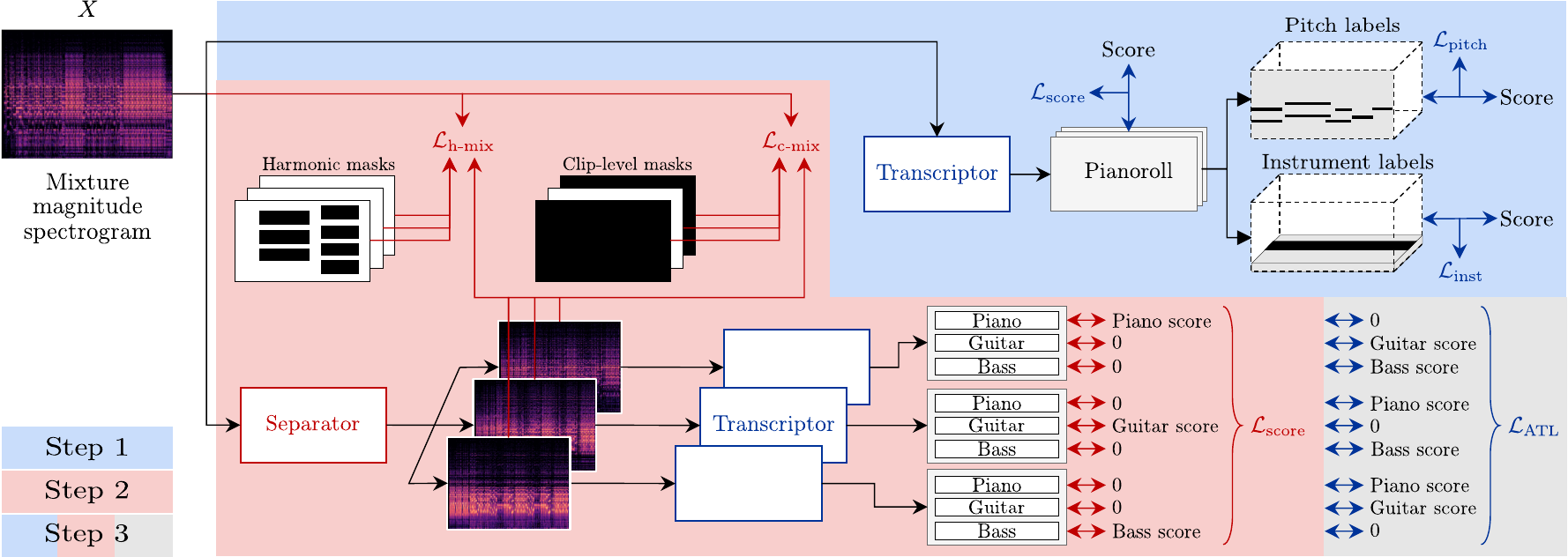}\vspace{-.2cm}
    \caption{Our proposed training strategy. Region colors indicate the associated training steps. %
    Red and blue loss terms are respectively used to update the separator and the transcriptor. The grey region shows the adversarial transcription loss described in Section~\ref{sec: step3}.}
    \label{fig: model}\vspace{-.3cm}
\end{figure*}

We adopt a common source separation pipeline. The input mixture is first transformed into a time-frequency (TF) representation $X=(X_{f,t})$ (e.g., STFT magnitude), where $f$ and $t$ denote time and frequency indices. A TF mask $M_{i}=(M_{i,f,t})$ for the $i$-th source (i.e., instrument) is then estimated by the source separation model from $X$. The separated TF representation $\hat{S}_{i}=(\hat{S}_{i,f,t})$ is obtained by multiplying $M_{i,f,t}$ with $X_{f, t}$ at each TF bin $(f,t)$. After combining each $\hat{S}_{i}$ with the mixture phase, an inverse transform (e.g., iSTFT) is applied to produce the estimated sources. %

During training, we assume only the music mixture and its score are available. Note that we here assume the audio and musical scores are well aligned, in practice using audio synthesized from MIDI. But as we only require scores during training (not at inference time), computationally intensive offline alignment algorithms such as those based on dynamic time warping~\cite{muller2015fundamentals} could be incorporated as a pre-processing step. We leave the alignment problem as future work.

Our proposed model is composed of two components, a \emph{ transcriptor} and a \emph{separator}. The model is trained using a three-step training approach. In the first step, we train a transcriptor to transcribe a mixture of music. Once the transcriptor is well trained, in the second step, we use the transcriptor as a critic to guide the learning of the separator. After this isolated training, the transcriptor and the separator are fine-tuned together in a final step. 

{\allowdisplaybreaks
\subsection{Step 1: Transcriptor Training} \label{sec: step1}
We first train the transcriptor to estimate the musical score given a mixture spectrogram, as illustrated in the blue region of Fig.~\ref{fig: model}. We represent the target score $Y=(Y_{i,n,t})$ as a multi-instrument piano roll, a tensor with instruments indexed by $i$, the 88 MIDI notes by $n$, and time by $t$. Following~\cite{hung2019multitask}, two marginalized representations, multi-pitch labels and instrument activations, are also derived from the piano roll as additional training targets to improve music transcription performance. The binary cross entropy loss $H(y, \hat{y}) = -y\log(\hat{y})-(1-y) \log(1-\hat{y})$ is used to evaluate the error between an estimated output probability $\hat{y}$ and a ground truth label $y$. %
The transcriptor is updated based on three loss terms, score estimation, instrument estimation, and multi-pitch estimation:
\begin{align}
    \mathcal{L}_{\operatorname{score}}(X,Y)&=\sum_{i,n,t} H(Y_{i,n,t}, p_{i,n,t}(X)),\label{bce-score}\\
    \mathcal{L}_{\operatorname{pitch}}(X,Y)&=\sum_{n,t} H(\max_i(Y_{i,n,t}), \max_i(p_{i,n,t}(X))),\label{bce-pitch} \\
    \mathcal{L}_{\operatorname{inst}}(X,Y)&=\sum_{i,t} H(\max_n(Y_{i,n,t}), \max_n(p_{i,n,t}(X))),\label{bce-inst}
\end{align}
where $p_{i,n,t}(X)$ denotes the estimated transcriptions for instrument $i$ at note $n$ and time $t$. %
The main transcription loss $\mathcal{L}_{\operatorname{TL}}$ is
\begin{align}\label{trans-loss-overall}
     \mathcal{L}_{\operatorname{TL}}&=\mathcal{L}_{\operatorname{score}}+\alpha_{1}\mathcal{L}_{\operatorname{inst}}+\beta_{1}\mathcal{L}_{\operatorname{pitch}}
\end{align}
where $\alpha_{1}$ and $\beta_{1}$ are weights used to balance the contribution of the marginalized objectives. %
}

\subsection{Step 2: Separator Training} \label{sec: step2}
We next use the transcriptor pre-trained in the first step with fixed parameters as a critic to train the separator, as described in the red region of Fig.~\ref{fig: model}. %
The separator has to output separated sources that are good enough for the transcriptor to transcribe into the correct score. The separator is given a mixture spectrogram and has to estimate separated spectrograms $\hat{S}_{i}$, which are then
given to the transcriptor as input for computing score estimates $p_{i,n,t}(\hat{S}_{i})$. Since each separated source should ideally only contain information from one instrument, the score estimated from the separated source should only have one active instrument while all others should be zero. The separator is updated via the sum of the transcription losses on each separated source $\hat{S}_i$:
\begin{align}\label{tran-loss}
    \mathcal{L}_{\operatorname{score}}(\hat{S}_i,Y^{(i)})\!=\! \sum_{n,t} \big( H(Y_i, p_{i,n,t}(\hat{S}_{i})) \!+\! \sum_{j\neq i} H(0, p_{j,n,t}(\hat{S}_{i}))\big),
\end{align}
where $Y^{(i)}$ are the labels obtained from $Y$ by setting to $0$ all sources other than $i$, i.e., $Y^{(i)}_{j,n,t}=\delta_{i,j} Y_{j,n,t}$.

A mixture loss is introduced in \cite{pishdadian2020finding} to prevent the output masks from solely focusing on the discriminating components for transcription, encouraging the sum of active components to be close to the input mixture, and the sum of inactive components to be close to zero.
Depending on whether ground truth activity is available at the clip or frame level, this leads to so-called clip-level and frame-level mixture losses.
As instrument activity tends to be consistent over the short clips (4 seconds) that we use for training, we here only consider the clip-level loss (frame-level loss lead to similar results): %
\begin{align}
  \mathcal{L}_{\operatorname{c-mix}} &= \sum_{f,t}|X_{f,t} - \sum_{i \in \mathcal{A}} \hat{S}_{i,f,t} | 
  +  \sum_{f,t} \sum_{i \notin \mathcal{A}} | \hat{S}_{i,f,t}|, \label{sep-loss-clip} 
\end{align}
where $\mathcal{A}$ is the index set of active sources in a training clip. %

\begin{figure}[t]
    \centering
    \includegraphics[width =
    1\columnwidth]{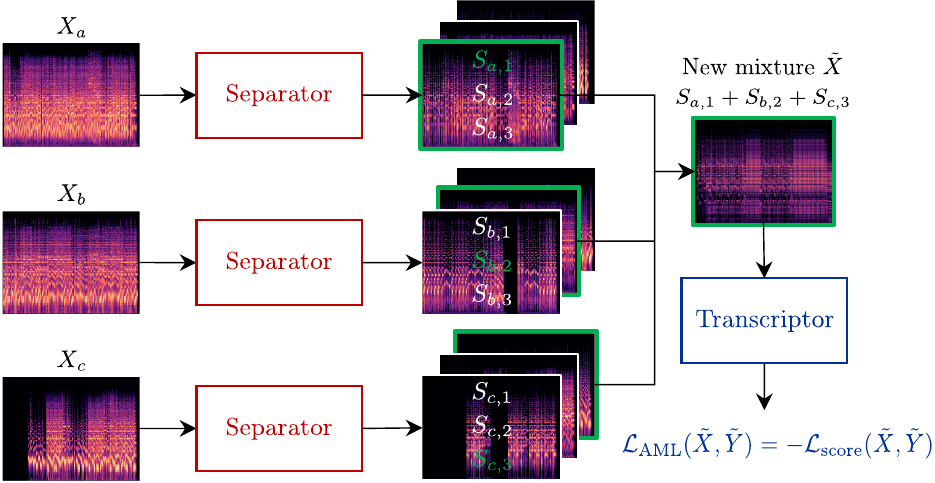}\vspace{-.2cm}
    \caption{Diagram of the adversarial mixture loss. The blue loss term is only used to update the transcriptor.}
    \label{fig: advMix}\vspace{-.1cm}
\end{figure}

To further leverage score information, we here introduce a harmonic mask mixture loss $\mathcal{L}_{\operatorname{h-mix}}$ to strengthen the harmonic structure of the separated sources. That is, we devise a within-frame notion of activity based on the harmonic structure of each instrument in each frame \cite{Miron2017MonauralSS}. %
The harmonic structure %
can be calculated from the piano roll representation. Assuming the tuning frequency of all songs is $f_q$ and its corresponding MIDI note number is $n_q$, we can compute the fundamental frequency for note $n$ as $f_0=f_q \cdot 2^{\frac{1}{12}\cdot(n - n_{q})}$. With the fundamental frequency, we can further calculate the harmonic frequencies $f_l = f_0 * (l + 1)$ where $l=1, \dots, L$ indexes the harmonics. We then create the set $\mathcal{A}_{f,t}$ of active sources expected to have energy in a given bin of the TF representation based on information from the score. This leads to the following loss:
\begin{align}\label{sep-loss-har}
  \mathcal{L}_{\operatorname{h-mix}} = \sum_{f,t}|X_{f,t} - \sum_{i \in \mathcal{A}_{f,t}} \hat{S}_{i,f,t} |
  +  \sum_{f,t} \sum_{i \notin \mathcal{A}_{f,t} } | \hat{S}_{i,f,t}|.
\end{align}
With this strong constraint, the model is able to focus on the harmonic structure. However, since the harmonic energy may not always concentrate at a single TF bin, and since fluctuations in $f_0$ may occur, frequencies within $\pm 1$ bin of an expected active harmonic $f_l$ are also considered active. %
Defining scalar weights $\alpha_{2}$ and $\beta_{2}$, the overall objective of separator training is 
\begin{align}\label{sep-loss-overall}
     \mathcal{L}_{\operatorname{sep}}&=\sum_{i}\mathcal{L}_{\operatorname{score}}(\hat{S}_i,Y^{(i)})+\alpha_{2}\mathcal{L}_{\operatorname{c-mix}}+\beta_{2}\mathcal{L}_{\operatorname{h-mix}}.
\end{align}

\subsection{Step 3: Joint Training} \label{sec: step3}
We now fine-tune the previously trained transcriptor and separator by training them together.
Without further loss terms, training of the transcriptor is not influenced by that of the separator, but we found that fine-tuning the separator against a slowly evolving transcriptor resulted in better performance than against a converged and fixed transcriptor.
We further introduce adversarial loss terms, as training the transcriptor to better transcribe the separated outputs would only cause it to co-adapt to the separator's mistakes, as shown in~\cite{pishdadian2020finding}. Instead, by forcing a competition between the transcriptor and the separator, the transcriptor should become more sensitive to errors in the separated outputs, instead of being robust to them.

We first consider an adversarial mixture loss $\mathcal{L}_{\operatorname{AML}}$, illustrated in 
Fig.~\ref{fig: advMix}. Denoting by $\hat{S}_{a,i}$ the $i$-th separated spectrogram for the $a$-th sample, we randomly pick one separated track for each instrument $i$ from a different sample $a_i$ to create a new remixed mixture $ \tilde{X} = \sum_{i} \hat{S}_{a_i,i}$, and corresponding remixed labels $\tilde{Y}$. Since the separated sources are not perfect, we ask the transcriptor to detect such fake mixtures from the real mixtures it is trained on, by encouraging it not to transcribe them properly via a negated loss:
\begin{align}\label{AML-loss}
    \mathcal{L}_{\operatorname{AML}}(\tilde{X},\tilde{Y}) = - \mathcal{L}_{\operatorname{score}}(\tilde{X},\tilde{Y}) 
\end{align}

We also consider an adversarial transcription loss
$\mathcal{L}_{\operatorname{ATL}}$ as a counterpart to~\eqref{tran-loss}, %
based on the assumption that the imperfect $\hat{S}_{i,f,t}$ contains incomplete information from the target instrument $i$ and residual information from other instruments. We thus ask the transcriptor not to recognize the target instrument, estimating zero as the target score, while transcribing non-target instruments to their correct scores,  leading to a loss function $\mathcal{L}_{\operatorname{ATL}}(\hat{S}_{i}, \overline{Y^{(i)}})= \mathcal{L}_{\operatorname{score}}(\hat{S}_i,\overline{Y^{(i)}})$, where $\overline{Y^{(i)}}=Y-Y^{(i)}$ is the complement of $Y^{(i)}$: %
\begin{align}\label{ALT-loss}
    \mathcal{L}_{\operatorname{ATL}}(\hat{S}_{i}, \overline{Y^{(i)}})= %
 \sum_{n,t} \big(   H(0, p_{i,n,t}(\hat{S}_{i})) + \sum_{j\neq i} H(Y_j, p_{j,n,t}(\hat{S}_{i})\big).
\end{align}
During step 3 training, the separator is still updated as in~\eqref{sep-loss-overall}, while the adversarial terms are added with weights $\alpha_{3}$ and $\beta_{3}$ to the original transcriptor loss~\eqref{trans-loss-overall}, resulting in 
\begin{align}
     \mathcal{L}_{\operatorname{TL-3}}&=\mathcal{L}_{\operatorname{TL}}+\alpha_{3}\mathcal{L}_{\operatorname{AML}}+\beta_{3}\mathcal{L}_{\operatorname{ATL}}.
\end{align}

\section{Experiment}

\subsection{Dataset}
Although many datasets have been proposed for music transcription, they have some limitations. For example, Bach10 \cite{Duan2010MultipleFF} and Su \cite{su2015escaping} contain a limited amount of songs; MusicNet \cite{thickstun2017learning} only includes classical music; MedleyDB \cite{bittner2014medleydb} has a comparably large amount of songs and genres, but only monopohonic instruments have score annotation, not polyphonic instruments such as guitars and pianos.
As a result, we leverage the recently released synthesized dataset Slakh2100 \cite{manilow2019slakh} for training and evaluating our system. Slakh is synthesized from the Lakh MIDI Dataset \cite{raffel2016learning} using professional-grade instruments. It contains aligned mono audio and MIDI score with 145 hours of data and 34 instrument categories. We use the \textit{Slakh2100-split2} training, validation, and test splits, and down-sample the audio to 16 kHz. We only consider mixtures of acoustic piano, distorted electric guitar, and electric bass in our current experiments, as we found certain synthesized instrument classes in Slakh2100 to be perceptually indistinguishable from one another (e.g., certain electric piano and clean electric guitar patches), and an inconsistent interpretation of MIDI drum scores used by the synthesis process.

\begin{table}
\caption{Separation performance in terms of SI-SDR (dB), where `isolated' and `fine-tune' respectively indicate the training setup of step 2 and step 3. %
}\vspace{0.1cm}
\label{tb:separation}
\centering
  \sisetup{table-format=2.1,round-mode=places,round-precision=1,table-number-alignment = center,detect-weight=true,detect-inline-weight=math}
\resizebox{\linewidth}{!}
  {\setlength{\tabcolsep}{2pt}
\begin{tabular}{lccccSSSS}
\toprule
  \multicolumn{1}{c}{Training}&\multicolumn{1}{c}{$\mathcal{L}_{\operatorname{c-mix}}$} &\multicolumn{1}{c}{$\mathcal{L}_{\operatorname{h-mix}}$} &\multicolumn{1}{c}{$\mathcal{L}_{\operatorname{AML}}$} &\multicolumn{1}{c}{$\mathcal{L}_{\operatorname{ATL}}$} &\multicolumn{1}{c}{Bass} &\multicolumn{1}{c}{Guitar} &\multicolumn{1}{c}{Piano} & \multicolumn{1}{c}{Avg}\\
\cmidrule(l{0.25em}r{0.25em}){1-1}\cmidrule(l{0.25em}r{0.25em}){2-3}\cmidrule(l{0.25em}r{0.25em}){4-5}\cmidrule(l{0.25em}r{0.25em}){6-9}
Supervised& & & & &
11.13 &5.66 &7.68 &8.16 \\
\midrule
isolated&\checkmark & & & &
7.48& 1.15& 4.19& 4.27 \\
isolated& &\checkmark & & &
7.82& 0.42& 4.14& 4.13 \\
isolated&\checkmark &\checkmark & & & \bfseries 8.40& \bfseries 1.60 & \bfseries 4.97 & \bfseries 4.99 \\
\midrule
fine-tune&\checkmark &\checkmark & & &8.96& 2.68& 5.27& 5.64 \\
fine-tune&\checkmark &\checkmark & \checkmark& & \bfseries 9.07 &\bfseries 2.80 &5.38 &\bfseries 5.75 \\
fine-tune&\checkmark &\checkmark & & \checkmark&
8.96 &2.51 &\bfseries 5.65 &5.71 \\
\specialrule{\heavyrulewidth}{0.4ex}{0.65ex}
Input mixture& & & & & 
1.15 &-5.83 &-2.34 &-2.34 \\
Baseline \cite{pishdadian2020finding}& & & & & 7.32& 0.49& 3.53& 3.78\\
\bottomrule
\end{tabular}
}\vspace{-.2cm}
\end{table}

\subsection{System Details} \label{sec: system}
We use the same temporal convolutional network (TCN) model for both our transcriptor and separator architectures since it has been shown to be efficient in source separation \cite{BaiTCN2018,luo2019convTasNet}. %
Our TCN contains three repeated layers with each layer including eight dilated residual blocks. All convolution layers have kernel size of 3. The output layers of the transcriptor and the separator are fully connected layers with sigmoid activation, the former producing a piano roll for each instrument, and the latter producing one TF mask per instrument.
The inputs to both transcriptor and separator are magnitude STFT. The STFT is computed using a square root Hann window of size 2048 samples and a hop size of 500. The output of the transcriptor is a three-dimensional piano-roll representation extracted using \texttt{pretty\_midi} \cite{raffel2014intuitive}. 
As described in \cite{pishdadian2020finding}, instrument dependent weights calculated based on the rate of frame-wise activity in the training set are applied to the transcription loss to compensate for class imbalance. The activity rate is 0.41 for piano, 0.14 for guitar and 0.67 for bass.
All the loss functions are given a weight to balance the loss during training process, with the weights manually tuned on development data and set to $\alpha_{1}=0.03$, $\beta_{1}=0.1$, $\alpha_{2}=1$, $\beta_{2}=0.1$, $\alpha_{3}=0.2$, and $\beta_{3}=0.05$. The various setups use subsets of the mixture and adversarial losses, and the weights of the unused terms are then set to $0$. %
Systems are trained with the Adam optimizer and a learning rate 0.001, which decays by a factor of 0.5 if the loss on the validation set does not improve for two consecutive epochs. Training stops if the validation loss does not improve for 10 epochs.

\subsection{Evaluation}
We compare our approach with the baseline method proposed in \cite{pishdadian2020finding}, which uses a classifier to perform weakly supervised source separation, with the difference that %
the classifier is designed to have the same TCN structure as the transcriptor for comparison. Only the last linear layer in the model is modified to output instrument activation. We report the scale-invariant signal-to-distortion (SI-SDR) ratio \cite{le2019sdr} of each instrument in various settings in Table \ref{tb:separation}. We observe that using a transcriptor leads to better separation performance than using a classifier. This shows that providing weak frequency information can benefit separating harmonic sounds, such as musical instruments. Although using a harmonic mask via  $\mathcal{L}_{\operatorname{h-mix}}$ alone has slightly lower performance, adding $\mathcal{L}_{\operatorname{h-mix}}$ with $\mathcal{L}_{\operatorname{c-mix}}$ %
improves average SI-SDR by 0.7 dB. This is likely because the harmonic mask is too strict in constraining the separated spectrogram, in particular forcing positions that are not exactly harmonic to be zero is not realistic and might influence separation quality. However, when combined with coarse clip-level activity constraints,
$\mathcal{L}_{\operatorname{h-mix}}$ can provide some information on harmonic structure which is lacking in $\mathcal{L}_{\operatorname{c-mix}}$. %

The bottom part of Table \ref{tb:separation} shows the result of jointly training the transcriptor and separator (fine-tune), starting from the best setup in step 2. We see that joint training without any adversarial loss improves separation performance, especially on guitar. As mentioned in Section~\ref{sec: step3}, this may be because, when fine-tuning the transcriptor and separator together, the slowly evolving transcriptor acts as a regularizer, preventing the separator from easily exploiting loopholes in the transcriptor. Moreover, by using an adversarial training strategy, we can further improve average SI-SDR by 0.2 dB, especially benefitting piano. We also include the supervised training result obtained with the same data split and separator architecture, where the separator is trained to minimize an $L_1$ loss function between $\hat{S}_i$ and the ground truth target spectrogram $S_i$. Other training settings are unchanged. Although the best weakly-supervised result is 2.4 dB worse than the supervised scenario, our approach still closes a significant part of the gap %
from the input SI-SDR (using mixture as input).

\begin{table}
\caption{Transcription performance in terms of note accuracy, where `pre-train' and `fine-tune' respectively indicate the training setup of step 1 and step 3, and %
`mixture' and `iso tracks' respectively represent mixture and isolated ground truth audio.}
\label{tb:transcription}\vspace{.1cm}
\centering
{\setlength{\tabcolsep}{4pt}
\begin{tabular}{lccccccc}
\toprule
 \multicolumn{1}{c}{Training}
 & \multicolumn{1}{c}{Evaluated on} 
 &\multicolumn{1}{c}{$\mathcal{L}_{\operatorname{AML}}$} &\multicolumn{1}{c}{$\mathcal{L}_{\operatorname{ATL}}$} &\multicolumn{1}{c}{Bass} &Guitar &Piano\\
\cmidrule(l{0.4em}r{0.4em}){1-1}\cmidrule(l{0.4em}r{0.4em}){2-2}\cmidrule(l{0.4em}r{0.4em}){3-4}\cmidrule(l{0.4em}r{0.4em}){5-7}
 pre-train & mixture & & &0.85 &0.44 &0.58\\ 
 fine-tune & mixture & & &0.84 &0.42 &0.54\\ 
 fine-tune & mixture &\checkmark& &0.86 &0.51 &0.61\\
 fine-tune & mixture & &\checkmark&0.85 &0.50 &0.60\\
 \midrule
 pre-train & iso tracks & & &0.91 &0.52 &0.66\\ 
 fine-tune & iso tracks & & &0.90 &0.53 &0.63\\ 
 fine-tune & iso tracks &\checkmark& &0.91 &0.58 &0.68\\
 fine-tune & iso tracks & &\checkmark&0.91 &0.57 &0.66\\
\bottomrule
\end{tabular}}
\vspace{-.2cm}
\end{table}

Table \ref{tb:transcription} compares the transcription results when first pre-training on mixtures alone (step 1) and when fine-tuning (step 3). Fine-tuning without adversarial loss simply means longer training, which may overfit. The note accuracy is calculated by \texttt{mir\_eval} \cite{raffel2014mir_eval}. The top section is evaluated using mixtures as the input. We observe that using adversarial training provides improvement in transcription accuracy, especially for guitar. Furthermore, we calculate the false positive rate of detected notes for all step 3 models in Table~\ref{tb:transcription} and find that the model trained without adversarial losses has a higher false positive rate (0.37) than models trained with $\mathcal{L}_{\operatorname{AML}}$ or $\mathcal{L}_{\operatorname{ATL}}$ (0.17 and 0.19 respectively). Because guitar and piano have similar note ranges, adversarial training appears to help the model distinguish between these two instruments. The lower section of Table \ref{tb:transcription} presents the results when evaluating on isolated ground truth audio. We observe a similar pattern as evaluating on the mixture, with guitar improving the most. This demonstrates that although our transcriptor is only trained on mixtures, it can generalize to isolated tracks.

\section{Conclusion}
In this work, we proposed a three-step training method to train a weakly-supervised music source separation model with musical scores. The results demonstrate that using a transcriptor to provide both temporal and frequency information as supervision outperforms the baseline method which only includes temporal information. We showed that using harmonic masks derived from the musical score, as well as adversarial losses on the transcriptor, leads to  improved separation and transcription performance. %
Future work includes tackling the alignment problem to expand the range of data that can be used for training, and considering separation of non-harmonic instruments such as drums within our framework. We also plan to explore semi-supervised approaches, combining small amounts of fully-supervised training data with large amounts of training data containing only mixtures and musical scores. 

\label{sec:conclusion}

\vfill\pagebreak

\bibliographystyle{IEEEtran}
\bibliography{refs}

\end{document}